\documentclass[conference]{IEEEtran}
\IEEEoverridecommandlockouts

\usepackage{balance}
\usepackage{cite}
\usepackage{xspace}
\usepackage{amsmath,amssymb,amsfonts}
\usepackage{multirow}
\usepackage{booktabs}
\usepackage{pifont}
\usepackage{graphicx}
\usepackage{colortbl}
\usepackage{array}
\usepackage{threeparttable}
\usepackage{url}

\usepackage{caption}
\usepackage{pgfplots}
\pgfplotsset{compat=1.16}

\usepackage{algorithm}
\usepackage{algpseudocode}

\usepackage{xcolor}
\usepackage{listings}
\definecolor{codegreen}{rgb}{0,0.6,0}
\definecolor{codegray}{rgb}{0.5,0.5,0.5}
\definecolor{codepurple}{rgb}{0.58,0,0.82}
\definecolor{backcolor}{rgb}{0.95,0.95,0.92}

\def\BibTeX{{\rm B\kern-.05em{\sc i\kern-.025em b}\kern-.08em
    T\kern-.1667em\lower.7ex\hbox{E}\kern-.125emX}}

\newcommand{\rblue}{\rowcolor{blue!10}}

\newcommand{\SystemName}{{\it ViM-Q}\xspace}

\newcommand{\eg}{{\it e.g.}\xspace}
\newcommand{\etal}{{\it et al.}\xspace}

\begin{document}

\title{ViM-Q: Scalable Algorithm-Hardware Co-Design for Vision Mamba Model Inference on FPGA}

\author{
    \IEEEauthorblockN{
        Shengzhe Lyu,
        Yuhan She,
        Patrick S. Y. Hung,
        Ray C. C. Cheung\IEEEauthorrefmark{1},
        Weitao Xu\IEEEauthorrefmark{1}
    }
    \IEEEauthorblockA{
        \textit{City University of Hong Kong, Hong Kong} \\
        \{shengzhe.lyu, yuhanshe3-c\}@my.cityu.edu.hk, \{psyhung, r.cheung, weitaoxu\}@cityu.edu.hk
    }
    \thanks{\IEEEauthorrefmark{1}Corresponding authors.}
}
\maketitle
\begin{abstract}
Vision Mamba (ViM) models offer a compelling efficiency advantage over Transformers by leveraging the linear complexity of State Space Models (SSMs), yet efficiently deploying them on FPGAs remains challenging. Linear layers struggle with dynamic activation outliers that render static quantization ineffective, while uniform quantization fails to capture the weight distribution at low bit-widths. Furthermore, while associative scan accelerates SSMs on GPUs, its memory access patterns are misaligned with the streaming dataflow required by FPGAs. To address these challenges, we present \textit{ViM-Q}~\footnote{Code available at: \url{https://github.com/shengzhelyu65/ViM-Q-FCCM-2026}}, a scalable algorithm-hardware co-design for end-to-end ViM inference on the edge. We introduce a hardware-aware quantization scheme combining dynamic per-token activation quantization and per-channel smoothing to mitigate outliers, alongside a custom 4-bit per-block Additive Power-of-Two (APoT) weight quantization. The models are deployed on a runtime-parameterizable FPGA accelerator featuring a linear engine employing a Lookup-Table (LUT) unit to replace multiplications with shift-add operations, and a fine-grained pipelined SSM engine that parallelizes the state dimension while preserving sequential recurrence. Crucially, the hardware supports runtime configuration, adapting to diverse dimensions and input resolutions across the ViM family. Implemented on an AMD ZCU102 FPGA, \textit{ViM-Q} achieves an average 4.96$\times$ speedup and 59.8$\times$ energy efficiency gain over a quantized NVIDIA RTX 3090 GPU baseline for low-batch inference on ViM-tiny. This co-design shows a viable path for deploying ViM models on resource-constrained edge devices.
\end{abstract}

\section{Introduction}
\label{sec:introduction}
Transformers~\cite{attention} have achieved remarkable success in computer vision tasks~\cite{vit, swin, deit, mae, dynamicvit, beit, taming, vitsurvey1, vitsurvey2}, yet the self-attention mechanism incurs a computational and memory complexity that is quadratic with respect to the input sequence length.
This inherent inefficiency poses significant challenges for deployment in resource-constrained or latency-sensitive environments.
State Space Models (SSMs), particularly Mamba~\cite{mamba, mamba2}, have emerged as a powerful alternative in various vision tasks~\cite{vim, vmamba, mambaout, mambavision, mobilemamba, localmamba, vimsurvey1, vimsurvey2, vimsurvey3}, offering linear-time complexity while maintaining or even surpassing the performance of Transformers.
Among these, Vision Mamba (ViM)~\cite{vim} successfully adapts this architecture for image tasks, demonstrating strong capabilities in visual representation learning.

\begin{figure}[t]
    \centering
    \includegraphics[width=0.98\linewidth]{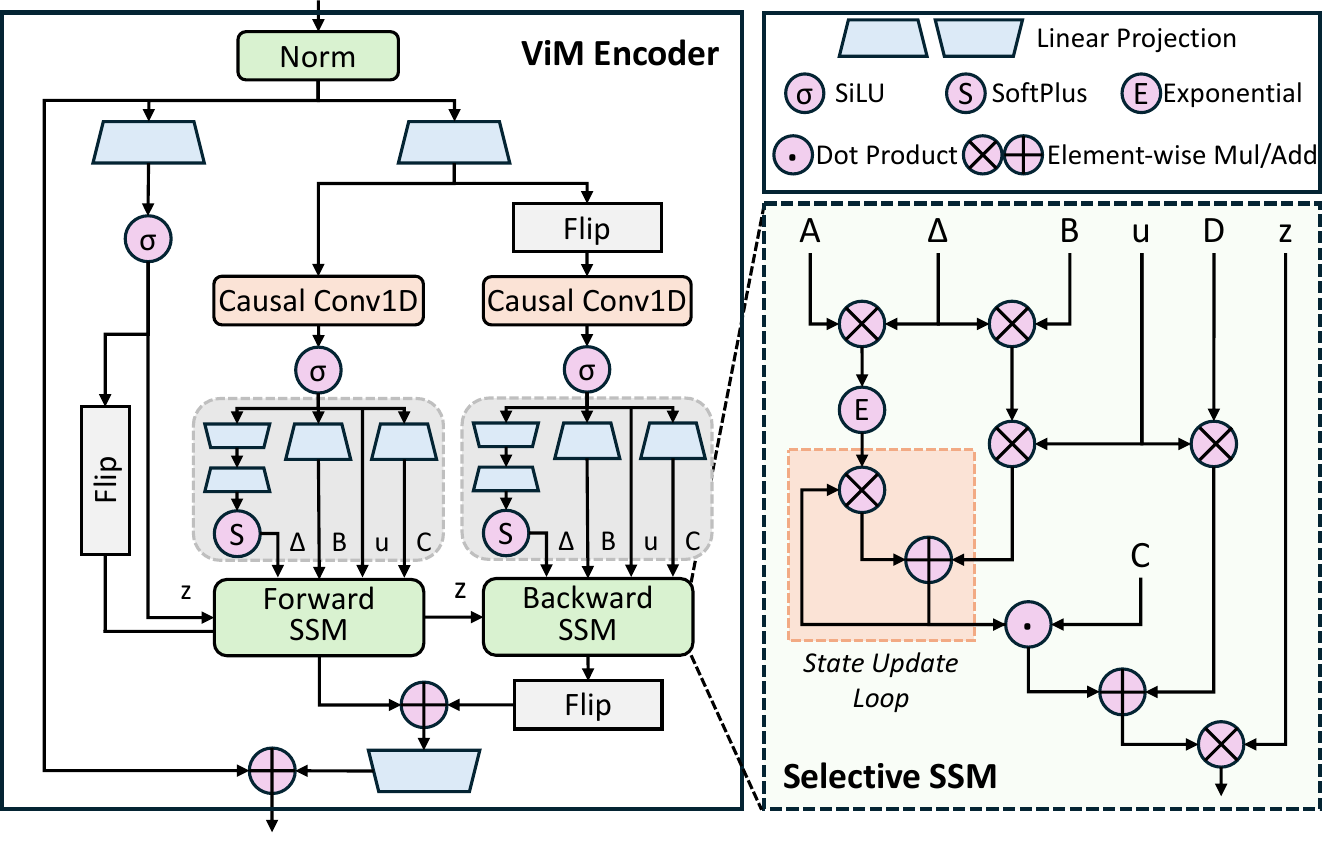}
    \captionsetup{skip=-2pt}
    \caption{Architecture of the ViM encoder and the detailed dataflow of the selective SSM mechanism.}
    \vspace{-5pt}
    \label{fig:introduction_architecture}
\end{figure}

Despite its promise, achieving efficient ViM deployment remains a significant research challenge.
Research focused on algorithms primarily targets quantization to reduce memory footprint and computational complexity.
While sophisticated Post-Training Quantization (PTQ) schemes~\cite{ptq4vm, mambaptq, mambaquant, quamba, quamba2, mambakscale, vimvq, ouromamba} excel at preserving accuracy, these methods frequently neglect hardware constraints.
They introduce complex operations that are inefficient for hardware mapping, which renders the quantized models impractical for edge implementations.

Similarly, deploying ViM on modern hardware platforms faces distinct hurdles.
General-purpose GPUs encounter bottlenecks in edge scenarios as they suffer from resource under-utilization at low batch sizes, which are typical workloads for real-time applications.
Furthermore, GPUs struggle to support low-bit quantization efficiently since they often rely on high-latency dequantization operations that undermine the theoretical benefits of reduced precision~\cite{lut,vsquant,ladder,tmac,awq}.
On the other hand, existing FPGA and ASIC accelerators~\cite{marca, lightmamba, mamba2fpga, mambax, specmamba, fastmamba} provide custom architectures but typically rely on rudimentary algorithmic optimizations such as static quantization. 
These approaches fail to accommodate the dynamic activation outliers inherent to ViM, which inevitably results in severe performance degradation.
This divergence creates a critical gap: hardware-centric solutions lack the necessary accuracy and adaptivity, while algorithmic solutions remain computationally impractical. A holistic algorithm-hardware co-design is therefore essential to unlock ViM's full potential on the edge.

Figure~\ref{fig:introduction_architecture} illustrates the ViM encoder architecture. The computational workload is dominated by dense linear projections of varying dimensions and the selective SSM, which acts as a content-aware gate.
Based on this architectural analysis, we identify three critical challenges in designing efficient hardware accelerators for ViM:

\textbf{(1) Hardware-aware quantization and efficient linear engine.}
Maintaining the accuracy of quantization is non-trivial for ViM since static methods fail on dynamic outliers, while uniform weight quantization is inefficient at low bit-widths.
Consequently, a dynamic activation scheme combined with a hardware-efficient weight representation is required.
To support this, the hardware needs a well-designed dataflow to hide the latency of dynamic quantization.
Furthermore, as resource constraints prohibit dedicated units for every linear transformation, a single unified linear engine is essential.

\textbf{(2) Mapping recurrent SSM to spatial parallelism.}
While GPU implementations leverage associative scan to accelerate SSM, migrating it to FPGAs requires prohibitively high memory bandwidth and storage requirements.
Thus, the challenge is to replace scans with a fine-grained dataflow that reconciles the sequential dependency of the SSM state update with high spatial parallelism.
Additionally, the dataflow must extend beyond the state update by fusing subsequent matrix and element-wise operations, thereby minimizing overall latency.

\textbf{(3) Scalability across model dimensions and resolutions.}
The ViM family encompasses diverse configurations characterized by varying hidden dimensions and arbitrary input resolutions.
To ensure deployment versatility, the objective is to achieve full runtime configuration across all compute modules, including linear, convolutional, and SSM engines.
This requires a specialized tiling strategy for each computation module, capable of adapting to arbitrary tensor shapes and sequence lengths without requiring hardware recompilation.

To address these challenges, we present \SystemName, to the best of our knowledge, \textit{the first end-to-end FPGA implementation of Vision Mamba models}, achieved through a scalable algorithm-hardware co-design. Our key contributions are:
\begin{itemize}
    \item We introduce a hardware-aware strategy that synergizes dynamic per-token activation quantization and per-channel smoothing with per-block 4-bit Additive Power-of-Two weight quantization, which effectively mitigates activation outliers to maintain high accuracy while maximizing the representational efficiency of low-bit weights.
    
    \item We propose a unified runtime-parameterizable linear engine employing a Lookup-Table computation unit to execute linear operations using pre-calculated results indexed by weights. This design replaces multiplications with shift-add operations in real-time, enabling efficient computation while hiding dynamic quantization latency.
    
    \item We design a pipelined SSM engine that exploits deep state-level parallelism while respecting the strict token-level recurrence of the state space update. By fusing subsequent matrix and element-wise operations into a unified streaming dataflow, this design significantly reduces intermediate memory access and overall latency.
\end{itemize}

We implement \SystemName on an AMD ZCU102 FPGA. 
Algorithmic evaluation confirms robust quantization performance, with ViM-t and ViM-s achieving 74.2\% and 79.9\% Top-1 accuracy at W4A8 precision, respectively. 
Hardware benchmarks demonstrate superior efficiency in low-batch scenarios, outperforming a quantized RTX 3090 GPU with an average 4.96$\times$ speedup and 59.8$\times$ energy efficiency gain for ViM-t.
We detail the proposed quantization scheme in Section~\ref{sec:sw_quant}, the hardware architecture in Section~\ref{sec:hardware_overview}, and optimizations for the linear and SSM engines are presented in Sections~\ref{sec:hw_linear} and~\ref{sec:hw_ssm}.
\section{Related Work}
\label{sec:related_work}

\subsection{Quantization of Mamba and Vision Mamba}
Mamba models, leveraging SSMs with linear time complexity, pose unique quantization challenges due to activation outliers.
Pierro~\etal~\cite{mambaptq} identified outlier channels and proposed selective quantization.
The Quamba framework~\cite{quamba} introduced 8-bit per-tensor quantization, extended by Quamba2~\cite{quamba2} with offline sorting, clustering, and per-state-group quantization.
For ViM, PTQ4VM~\cite{ptq4vm} uses per-token static quantization and joint learning of smoothing scales, while Shi~\etal~\cite{mambakscale} apply k-scaled token-wise quantization and hidden state reparameterization.
Deng~\etal propose ViM-VQ~\cite{vimvq}, which employs post-training vector quantization with optimized codewords.
OuroMamba~\cite{ouromamba} uses data-free quantization with synthetic data, and MambaQuant~\cite{mambaquant} applies smooth-fused rotation to address uneven data distributions.

\subsection{Hardware Accelerators for Mamba}
Hardware accelerators enhance Mamba model performance, particularly on FPGAs.
LightMamba~\cite{lightmamba} uses hardware-software co-design with up to 4-bit quantization, partial unrolling, and optimized tiling.
Zhou~\etal's accelerator~\cite{mamba2fpga} for Mamba-2 employs a reconfigurable architecture with an intra-layer pipeline for SSM throughput.
Li~\etal's Marca~\cite{marca} features a PE array for linear and nonlinear operations, with advanced buffer management.
Yoon~\etal propose Mamba-X~\cite{mambax}, which targets ViM on edge devices using a systolic scan array and hybrid quantization.
SpecMamba~\cite{specmamba} introduces a speculative execution framework that accelerates inference by predicting future tokens and verifying them in parallel on an FPGA, while FastMamba~\cite{fastmamba} presents a dedicated FPGA accelerator to optimize SSM computations through efficient memory access and pipelining.

As summarized in Table~\ref{tab:related_work_comparison}, our work pioneers the FPGA implementation of Vision Mamba, employing a novel W4A8 dynamic PTQ scheme. 
This distinguishes our design from prior accelerators that rely on static quantization~\cite{lightmamba, mamba2fpga}, target standard Mamba~\cite{lightmamba, mamba2fpga, marca}, or utilize ASICs~\cite{marca, mambax}.
\section{Hardware-Aware Quantization}
\label{sec:sw_quant}

\begin{table}[t!]
\centering
\caption{Qualitative comparison with other hardware accelerators for Mamba-based models.}
\label{tab:related_work_comparison}
\vspace{-5pt}
\resizebox{\columnwidth}{!}{%
\begin{tabular}{@{}l|ccccc@{}}
\toprule
\textbf{Feature} & ~\cite{lightmamba} & ~\cite{mamba2fpga} & ~\cite{marca} & ~\cite{mambax} & \textbf{Ours} \\
\midrule
\textbf{Target Model} & Mamba & Mamba-2 & Mamba & Vision Mamba & \textbf{Vision Mamba} \\
\textbf{Target Platform} & FPGA & FPGA & ASIC & ASIC & \textbf{FPGA} \\
\textbf{Bit Precision} & W4A4/W8A8 & W8A8/W8A16 & FIX32 & W8A8 & \textbf{W4A8} \\
\textbf{Quantization} & Static PTQ & Static PTQ & None & Static PTQ & \textbf{Dynamic PTQ} \\
\textbf{SSM Dataflow} & Unrolled & Pipeline & Tree-Based & Systolic Scan & \textbf{Spatial Recurrent} \\
\bottomrule
\end{tabular}%
}
\vspace{-5pt}
\end{table}
\begin{figure}[t!]
    \centering
    \includegraphics[width=0.9\linewidth]{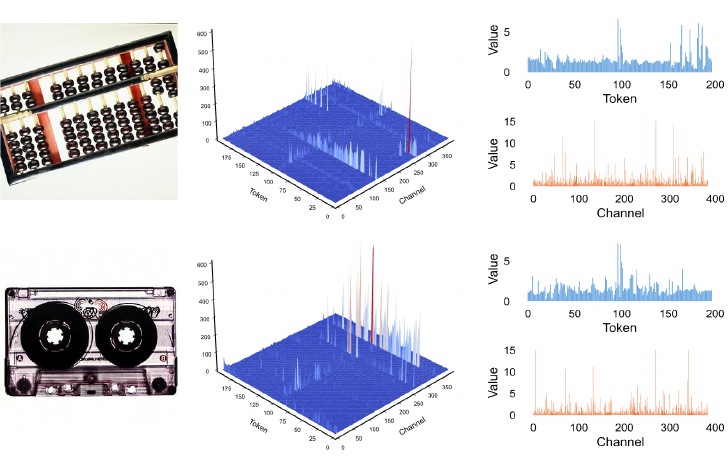}
    \captionsetup{skip=-2pt}
    \caption{Input images and the corresponding input activation distributions of the output projections at the 22nd layer in the tiny ViM model variant.}
    \vspace{-5pt}
    \label{fig:background_activation_weights}
\end{figure}

To enable efficient FPGA deployment, we target the quantization of linear and convolutional layers. 
Consistent with prior art~\cite{mamba2fpga, mambax, ptq4vm, mambaptq, mambaquant, mambakscale, vimvq, ouromamba}, we retain the SSM module in high precision to avoid severe error accumulation from recurrent state updates.
Activation quantization faces dual challenges. 
First, similar to other large vision models, ViM exhibits persistent channel-wise outliers driven by weight distributions~\cite{mambaptq, mambaquant, ptq4vm}.
Second, the model generates highly dynamic, input-dependent per-token outliers~\cite{ouromamba, mambaquant, ptq4vm, quamba, quamba2}, as shown in Figure~\ref{fig:background_activation_weights}.
This compound variability renders standard static quantization ineffective.
Regarding weights, the distribution is typically Gaussian, making uniform quantization inefficient as it wastes quantization levels on sparse tails.
Consequently, a non-uniform scheme is essential to minimize bit-width without compromising accuracy.

\subsection{Per-Channel Activation Smoothing}
We quantize activations to INT8 to exploit FPGA-native integer arithmetic.
To mitigate channel-wise outliers, we adopt smoothing~\cite{smoothquant, svdquant} to redistribute quantization difficulty from activations to weights via scaling factors \(s_j = \max(|X_j|)^\alpha/\max(|W_j|)^{1-\alpha}\), where $X_j$ denotes the $j$-th activation channel, $W_j$ denotes the corresponding weights, and $\alpha = 0.5$ is the balancing hyperparameter.
To minimize runtime overhead, we fuse the smoothing operation directly into the weights.
Smoothing is arithmetically equivalent to scaling down the upstream weights while proportionally scaling up the downstream weights.
This allows for offline transformation, requiring the explicit insertion of a smoothing layer only when nonlinearities interrupt linear layers.

\subsection{Dynamic Per-Token Quantization}
While smoothing alleviates channel outliers, static activation quantization remains suboptimal.
A fixed scale derived from a calibration set must guard against rare extremes, resulting in conservative scaling that underutilizes the INT8 range and risks clipping on unseen inputs.
To address this, \SystemName implements dynamic per-token quantization.
For each token, a dedicated unit calculates the real-time absolute maximum to derive an optimal scaling factor.
This maximizes dynamic range utilization, preserving information that is otherwise lost under static activation quantization schemes.

\subsection{Per-Block Additive Power-of-Two Weight Quantization}
\lstset{
    backgroundcolor=\color{backcolor},
    basicstyle=\scriptsize\ttfamily,
    numbers=left,
    numberstyle=\tiny\color{codegray},
    keywordstyle=\color{blue},
    commentstyle=\color{codegreen},
    stringstyle=\color{codepurple},
    showstringspaces=false,
    breaklines=true,
    frame=single,
    framerule=1pt,
    rulecolor=\color{codegray},
    tabsize=2,
    language=Python,
    escapechar=!
}

\begin{figure}[t]
  \centering
  \begin{minipage}{0.95\linewidth}
    \begin{lstlisting}[language=Python]
def PerBlock_APoT_WeightQuant(w, B, Q):
    """
    w : weights, B : block size, Q : APoT levels
    """
    # 1. Reshape tensor into blocks
    w_blk = !\textcolor{blue}{reshape}!(w, (-1, B))
    # 2. Compute Per-Block Scale
    s = !\textcolor{blue}{max}!(!\textcolor{blue}{abs}!(w_blk), axis=1)
    # 3. Normalize and Clip
    w_n = !\textcolor{blue}{clamp}!(w_blk / s, -1, 1)
    # 4. Extract Sign and Magnitude
    sign = !\textcolor{blue}{sign}!(w_n)
    mag  = !\textcolor{blue}{abs}!(w_n)
    # 5. Map Magnitude to nearest APoT level
    idx = !\textcolor{blue}{argmin}!(|mag - Q|)
    val = Q[idx]
    return val, sign, s
    \end{lstlisting}
  \end{minipage}
  \captionsetup{skip=-1pt}
  \caption{Post-training per-block APoT weight quantization.}
  \vspace{-3pt}
  \label{fig:software_apot_quant}
\end{figure}

We replace standard uniform weight quantization with an aggressive 4-bit per-block Additive Power-of-Two (APoT) scheme.
APoT restricts quantization levels to sums of power-of-two terms, enabling non-uniform distribution that allocates higher resolution near zero, where most weights reside~\cite{mixmatch, p2vit, li2020Additive, yao2022rapq, przewlocka2022power}.
Crucially, this structure allows hardware to replace costly multiplications with logic-efficient bit-shift operations.

\textbf{Per-Block Granularity.}
A major challenge in low-bit quantization is weight outliers, which skew dynamic ranges and degrade precision. Per-channel scaling is often too coarse to mitigate this.
To address this, we implement fine-grained per-block quantization as illustrated in Figure~\ref{fig:software_apot_quant}.
By reshaping the weight tensor into small, independent blocks with dedicated scaling factors, we ensure outlier influence is isolated to local regions, significantly preserving fidelity~\cite{vsquant, zeroquant, gpt3}.

\textbf{4-bit APoT Construction.}
We utilize a 4-bit format comprising 1 sign bit and 3 magnitude bits. To optimize representational capacity, the magnitude indexes 8 levels constructed via a split-basis approach: a 2-bit Coarse Basis $b_C$ and a 1-bit Fine Basis $b_F$.
As detailed in Table~\ref{tab:basis_b4}, the final magnitude $val = c + f$ is the sum of one element from each group.
This hierarchical construction maintains high accuracy by effectively covering the weight distribution while dramatically reducing the memory footprint.

\begin{table}[t]
\centering
\caption{Basis components for 4-bit APoT weights. The final 8 quantization levels are formed by the set $\{c+f | c \in b_C, f \in b_F\}$.}
\renewcommand{\arraystretch}{1.1}
\vspace{-5pt}
\resizebox{0.85\columnwidth}{!}{
\begin{tabular}{c|c|c}
\hline
\textbf{Basis Group} & \textbf{Exponent Values ($k$)} & \textbf{Numerical Values} \\
\hline
Coarse Basis ($b_C$) & $\{1, 2, 4\}$ & $\{0, 2^{-1}, 2^{-2}, 2^{-4}\}$ \\
Fine Basis ($b_F$) & $\{3\}$ & $\{0, 2^{-3}\}$ \\
\hline
\end{tabular}
}
\vspace{-5pt}
\label{tab:basis_b4}
\end{table}

\section{Hardware Design Overview}
\label{sec:hardware_overview}

\begin{figure*}[t]
    \centering
    \vspace{-5pt}
    \includegraphics[width=0.95\linewidth]{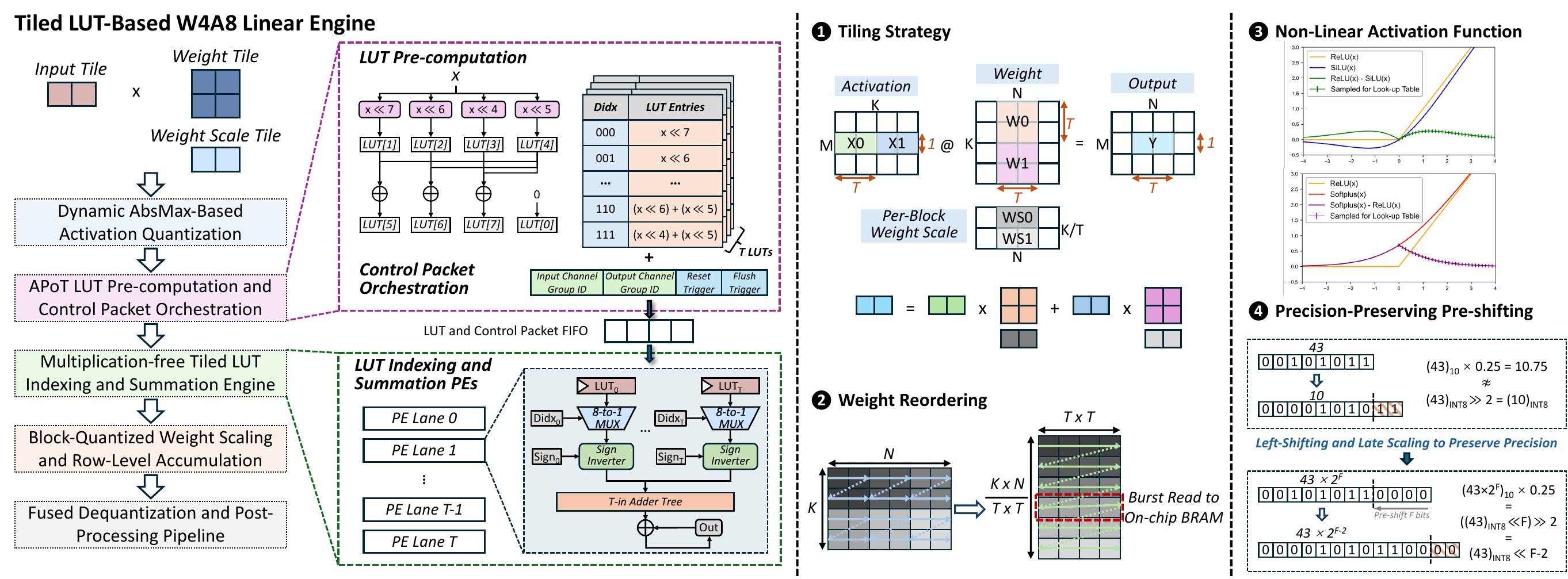}
    \captionsetup{skip=0pt}
    \caption{Unified linear engine design with LUT pre-computation. Key optimizations include: \ding{202} streaming tiling strategy, \ding{203} memory-efficient weight reordering, \ding{204} LUT-based non-linear activations, and \ding{205} precision-preserving pre-shifting.}
    \vspace{-13pt}
    \label{fig:hardware_linear}
\end{figure*}

The \SystemName architecture follows a modular, streaming design where specific ViM functions are implemented as standalone engines.
Each engine integrates local BRAM buffers and dedicated burst loaders to maximize AXI bandwidth, while supporting runtime configuration via control parameters (\eg, feature dimensions and patch counts).
Operating on a tiling-based dataflow, engines process partitioned input tiles immediately upon arrival, communicating via lightweight FIFO buffers.
This decoupled pipeline facilitates concurrent execution across modules, effectively hiding memory latency to sustain high throughput.
The accelerator comprises three primary compute engines:

\textbf{Unified Linear Engine.}
Designed to handle all linear transformations, this engine features a datapath that integrates a dynamic activation quantizer, a Lookup-Table (LUT) pre-computation unit for generating bit-shifted tile results, and a high-throughput LUT indexing and summation core. The pipeline concludes with a post-processing unit supporting dequantization, fused bias addition, and activation functions.

\textbf{Fine-Grained Pipelined SSM Engine.}
Implemented as a dedicated three-stage pipeline, this engine orchestrates state space updates, state projection, and output generation. It harmonizes the SSM's sequential recurrence with spatial parallelism: while temporal dependencies are resolved via a low-latency streaming pipeline, the independent feature and state computations are distributed across parallel hardware lanes to maximize throughput and scalability.

\textbf{Auxiliary Engines.}
Lightweight modules handle essential operations including patch embedding, causal convolution, normalization, residual summation, and patch manipulation (\eg, CLS token extraction and sequence flipping).
Each engine is fully dataflow-compatible to minimize stalls. Notably, the causal convolution engine optimizes latency by decomposing windowing and filtering into distinct sub-operations.
Like the linear engine, it leverages dynamic quantization and bit-shift computation to ensure efficient execution.
\section{Unified Linear Engine Design}
\label{sec:hw_linear}

\begin{figure}[t]
    \centering
    \includegraphics[width=0.62\linewidth]{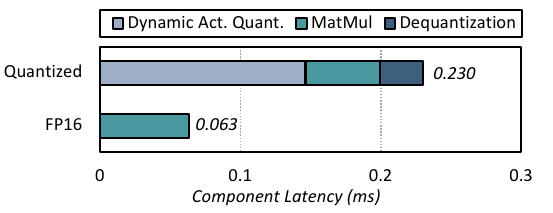}
    \captionsetup{skip=0pt}
    \caption{Latency breakdown of quantized versus FP16 linear layers (In=384, Out=768) on an NVIDIA RTX 3090.}
    \vspace{-5pt}
    \label{fig:hw_linear_gpu_breakdown}
\end{figure}

Dynamic quantization introduces a severe runtime bottleneck on general-purpose hardware. 
As illustrated in Figure~\ref{fig:hw_linear_gpu_breakdown}, the overhead of computing activation statistics and performing dequantization on the GPU consumes significantly more time than the matrix multiplication itself.
While custom FPGA architectures offer a promising solution, unlocking their full efficiency requires addressing three critical challenges:

\textbf{First}, to eliminate the dynamic quantization overhead, the engine necessitates a specialized streaming dataflow capable of computing quantization parameters on-the-fly, effectively hiding the latency within the execution pipeline.
\textbf{Second}, leveraging the benefits of our 4-bit APoT scheme requires a departure from standard arithmetic units.
Given the properties of power-of-two values, the challenge is to design an architecture that converts APoT arithmetic into logic-efficient bit-shifting operations.
\textbf{Third}, maximizing throughput across the diverse layer shapes of ViM variants necessitates a flexible tiling strategy to support full runtime reconfigurability.
Figure~\ref{fig:hardware_linear} illustrates our unified engine workflow, which addresses these challenges through a LUT-based datapath and four critical hardware optimizations (\ding{202}–\ding{205}).

\subsection{Tiling-Based Dataflow Architecture}
To maximize generality, all linear layers map to a unified engine, consistent with prior art~\cite{marca, lightmamba, edgemoe, finn, famous}.
Operations are decomposed into fixed-size tiles, as shown in Figure~\ref{fig:hardware_linear}-\ding{202}.
Input activations stream as $1 \times T$ vectors and are multiplied against $T \times T$ weight tiles from local BRAMs, while associated per-block scales are retrieved to rescale intermediate results. Partial sums accumulate until a full output tile is produced.
To align memory access, we employ offline weight reordering as depicted in Figure~\ref{fig:hardware_linear}-\ding{203}.
The standard matrix is transformed into a hardware-friendly layout where quantized weights are packed into sequential wide words (\eg, matching the 256-bit AXI burst length). This eliminates irregular indexing, enabling purely sequential burst reads to minimize DRAM latency.

\subsection{Dynamic Quantizer and Dequantizer}
Dynamic quantization is critical for model robustness but requires efficient hardware to avoid overhead.
Our architecture integrates a streaming quantizer that operates on-the-fly.
As activation tiles stream in, a compute unit calculates the maximum absolute value via parallel reductions.
Each token's global maximum is converted into a scaling factor, buffered in a FIFO, and forwarded to a division unit where activations are mapped to 8-bit integers. Simultaneously, these scales propagate downstream to the dequantizer for reconstruction.

The dequantizer reverses this process by rescaling intermediate results from the linear core. Fused operations like bias addition and nonlinearities are conditionally applied within the same datapath. To optimize complex activations, we employ a LUT-based approximation as shown in Figure~\ref{fig:hardware_linear}-\ding{204}, leveraging the fact that functions like SiLU and SoftPlus differ from ReLU by only a small, symmetric offset.
Thus, activations pass through a ReLU, requiring no computation other than thresholding at zero, and only the difference is retrieved from the LUT. Exploiting symmetry allows us to store only half the LUT entries, reducing on-chip memory usage.

\begin{figure}[t]
    \centering
    \includegraphics[width=0.7\linewidth]{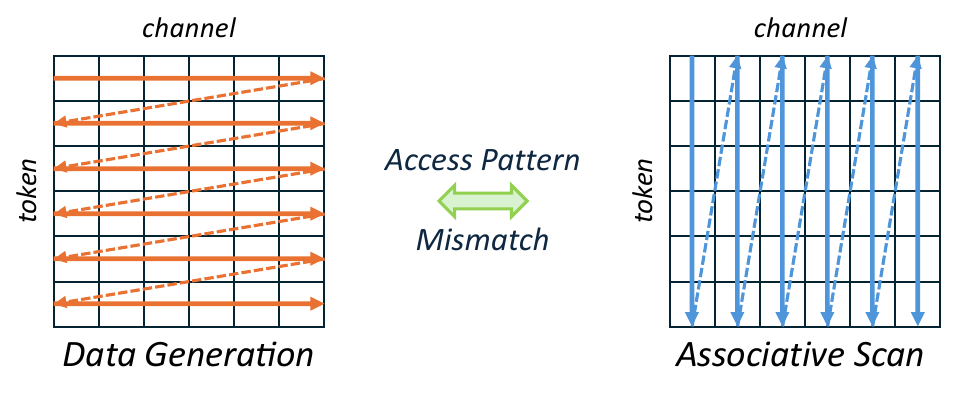}
    \captionsetup{skip=0pt}
    \caption{Data access misalignment between the token-major input stream and the channel-major traversal of the associative scan.}
    \vspace{-6pt}
    \label{fig:hardware_scan_mismatch}
\end{figure}

\subsection{LUT-Based Linear Computation}
While APoT theoretically replaces multiplications with bit-shifts, a naive implementation via variable barrel shifters per Processing Element (PE) is inefficient.
Since our 4-bit weights (3-bit magnitude, 1-bit sign) map to only eight shifting patterns, placing dedicated shifters in every PE results in redundant recalculation and excessive logic overhead.
We propose a LUT-based architecture that decouples shifting from the computation loop, transforming arithmetic into lightweight indexing and addition~\cite{lut, tmac, lutdla, figlut, lutgemm}.

\textbf{LUT-Based Pre-computation.}
Instead of dynamic shifting within each PE, the engine pre-calculates shifted results for each input tile only once.
As a $1 \times T$ activation tile streams in, a unit generates eight potential results per element without considering the sign, populating $T$ parallel LUTs.
To ensure downstream PEs remain stateless, a control packet containing tile coordinates (\eg, group IDs) and sync flags (reset/flush) propagates alongside the LUT data, allowing subsequent PEs to manage partial sum accumulation and result flushing.

\textbf{Precision-Preserving Pre-shifting.}
As detailed in Figure~\ref{fig:hardware_linear}-\ding{205}, direct integer right-shifting discards fractional bits, leading to accuracy degradation.
We mitigate this by left-shifting inputs by $F$ bits (\eg, $F=8$) to widen the dynamic range.
To minimize overhead, we fuse this into LUT generation: for weight basis $2^{-k}$, the engine directly computes $x \ll (F - k)$.
This preserves fractional precision throughout accumulation, with the final result right-shifted by $F$ to compensate for the pre-shifting and restore the numerical range.

\textbf{LUT Indexing and Summation PEs.}
As shown in Figure~\ref{fig:hardware_linear}, the core linear operation is subsequently reduced to a streamlined indexing and summation process, where pre-computed LUTs are broadcast to $T$ parallel PE lanes, each corresponding to an output channel.
Inside each lane, the 3-bit weight magnitude index controls an 8-to-1 multiplexer to select the pre-computed term from the LUT, while the sign bit drives a conditional inverter for negation.
These retrieved values are then summed via an adder tree.
By resolving the block-level accumulation locally, the PE outputs a single aggregated result for a weight block, which is then passed to the subsequent stage for scaling and row-level accumulation before final dequantization.
\section{Fine-Grained Pipelined SSM Design}
\label{sec:hw_ssm}

The selective SSM challenges efficient ViM inference due to its sequential state updates.
GPUs accelerate it via associative scan~\cite{mamba, mamba2, vim, scan}, but this conflicts with FPGA's streaming model.
As shown in Figure~\ref{fig:hardware_scan_mismatch}, a data access conflict arises: input arrives in token-major order, while associative scan requires channel-major traversal.
Resolving this mismatch demands an expensive on-chip transposition buffer, consuming excessive BRAM and increasing control complexity.
Thus, efficiently mapping the selective SSM onto spatially parallel FPGA architectures remains a key design challenge.

\subsection{Pipelined SSM Architecture and Tiling}
\begin{figure}[t]
    \centering
    \includegraphics[width=\linewidth]{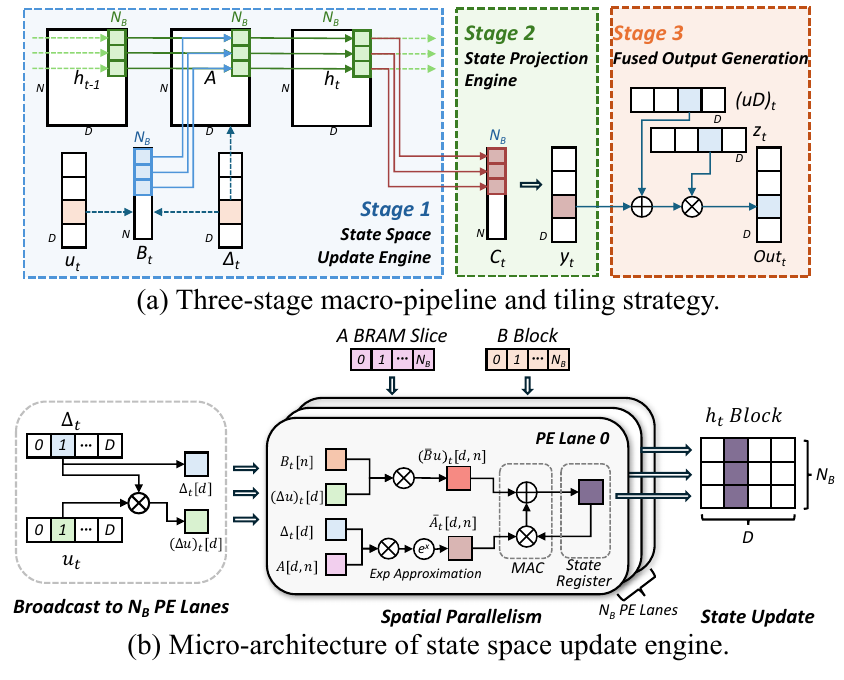}
    \captionsetup{skip=-2pt}
    \caption{Fine-grained pipelined SSM architecture.}
    \vspace{-5pt}
    \label{fig:hardware_ssm}
\end{figure}

To exploit FPGA spatial parallelism, we design a fine-grained accelerator that decomposes SSM computation into three decoupled, concurrently operating macro-pipeline stages: \textit{State Space Update}, \textit{State Projection}, and \textit{Fused Output Generation}, as shown in Figure~\ref{fig:hardware_ssm}(a).
We employ a tiling strategy along the state dimension, processing blocks of $N_B$ states in parallel to adapt to varying state dimensions.

\subsection{Stage 1: State Space Update Engine}
This stage integrates discretization and state space recurrence. It streams inputs $\Delta, u \in \mathbb{R}^{L \times D}$ and time-variant projection $B \in \mathbb{R}^{L \times N}$ from off-chip memory, while the time-invariant transition matrix $A \in \mathbb{R}^{D \times N}$ resides in on-chip BRAM.
Here, $L$, $D$, and $N$ denote the sequence length, feature dimension, and state dimension, respectively.

\textbf{Discretization Step.}
We employ a broadcasting architecture to compute the discretized state transition gate $\bar{A} = \exp(\Delta \cdot A)$ and the input-modulated control term $\bar{B}u = (\Delta \cdot u) \cdot B$, as illustrated in Figure~\ref{fig:hardware_ssm}(b).
For token $t$ and channel $d$, the scalar pair $(\Delta_{t}[d], u_{t}[d])$ is fetched. The product $(\Delta u)_{t}[d]$ is computed locally, and both $\Delta_{t}[d]$ and $(\Delta u)_{t}[d]$ are broadcast across $N_B$ parallel PE lanes. These lanes simultaneously receive the projection slice $B_t$ and the corresponding local weights $A$.
Inside the PEs, two parallel operations execute:
(1) The $\Delta$ value multiplies local $A$ weights via an optimized exponential approximation to generate the decay gate $\bar{A}$.
(2) The $\Delta u$ value multiplies vector $B_t$ to produce $\bar{B}u$.
This dataflow decouples the feature dimension $D$ from the state expansion $N$, enabling high spatial parallelism.

\textbf{Iterative State Update.}
Following discretization, the engine performs the core recurrence update:
\setlength{\abovedisplayskip}{3pt}
\setlength{\belowdisplayskip}{3pt}
\begin{equation}
    h_t = h_{t-1} \odot \bar{A} + \bar{B}u.
\end{equation}
Hidden states $h_t$ reside entirely in distributed register files.
The calculated $\bar{B}u$ is fused directly into a Multiply-Accumulate (MAC) operation with the previous state $h_{t-1}$ and $\bar{A}$. This ensures state updates occur with single-cycle latency, producing the updated state tensor $h \in \mathbb{R}^{L \times D \times N}$ for the next stage.

\subsection{Stage 2: State Projection Engine}
This stage functions as a streaming reduction unit and maps the high-dimensional hidden state back to the feature space:
\setlength{\abovedisplayskip}{3pt}
\setlength{\belowdisplayskip}{3pt}
\begin{equation}
    y_t = h_t \cdot C_t,
\end{equation}
where $C_t \in \mathbb{R}^{N}$ is the token-dependent output weight vector streamed from off-chip memory.
The updated state blocks are multiplied by the corresponding slices of $C_t$. The results are then passed through an adder tree to accumulate values across the state dimension, compressing the $N$-dimensional state into a single scalar output $y_t[d]$ for each feature channel.

\subsection{Stage 3: Fused Output Generation}
The final stage generates the module output by applying the residual connection and gating mechanism.
To minimize memory traffic and latency, we fuse these element-wise operations into a single combinational pipeline.
The unit accepts the projected features $y_t$ from Stage 2 and simultaneously streams the original input token $u_t$, the skip parameter $D$, and the gating factor $z_t$ from memory.
It executes the following composite operation in a single pass:
\setlength{\abovedisplayskip}{3pt}
\begin{equation}
    Out_t = (y_t + u_t \odot D) \odot z_t.
\end{equation}
\section{Experiments}
\label{sec:evaluation}

\subsection{Experimental Setup}
\textbf{Algorithm.}
We evaluate our proposed quantization methodology on the tiny (ViM-t), small (ViM-s), and base (ViM-b) ViM model variants, as detailed in Table~\ref{tab:evaluation_sw_model_configs}.
We assessed their performance on the ImageNet-1k classification benchmark, reporting both Top-1 and Top-5 accuracy.

\begin{table}[t]
\centering
\caption{Vision Mamba model configurations.}
\label{tab:evaluation_sw_model_configs}
\footnotesize
\vspace{-6pt}
\resizebox{0.6\columnwidth}{!}{%
\begin{tabular}{lccc}
\toprule
\textbf{Attribute} & \textbf{Tiny} & \textbf{Small} & \textbf{Base} \\
\midrule
Model           & ViM-t & ViM-s & ViM-b \\
Parameters      & 7 M   & 26 M  & 98 M \\
\# Encoder blocks & 24    & 24    & 24   \\
Hidden dimension & 192   & 384   & 768  \\
State dimension  & 16    & 16    & 16   \\
\bottomrule
\end{tabular}
}
\vspace{-5pt}
\end{table}

\textbf{Hardware Implementation.}
The proposed accelerator was implemented on an AMD ZCU102 evaluation board.
Individual hardware modules were synthesized using C++-based Vitis HLS 2025.2.
These modules were integrated into the top-level end-to-end design using SpinalHDL 1.10.1, which generated the system Verilog RTL.
The final logic synthesis, place-and-route, and bitstream generation were completed using Vivado 2025.2.
On-chip power consumption estimates were derived from the post-implementation reports generated by Vivado.

\textbf{Baselines and Measurement.}
We benchmark \SystemName against an NVIDIA RTX 3090 GPU using FP16 and quantized baselines at a batch size of 1.
Both baselines utilize fused CUDA kernels for the parallel associative scan to accelerate SSM recurrence.
Lacking native APoT arithmetic support, the quantized GPU baseline uses a dequantization-based approach: compressed weights are dynamically cast back to FP16 before execution on standard tensor cores.
Power and latency are recorded via the NVIDIA Management Library (NVML) and averaged over ten iterations after a warm-up phase.
Active chip power is isolated by subtracting an estimated 50\,W idle board and fan power from raw measurements.
\subsection{Evaluation of Quantization Scheme}
\label{sec:eval_quant}

\subsubsection{Performance of Quantization Scheme}

\begin{table}[t]
\centering
\caption{Evaluation of ViM models across different quantization precisions and granularities (per-channel vs. per-block).}
\label{tab:evaluation_sw_quantization}
\vspace{-5pt}
\resizebox{0.95\columnwidth}{!}{%
\begin{tabular}{lcccc}
\toprule
\textbf{Methods} & \textbf{W/A Precision} & \textbf{Granularity} & \textbf{Top 1 (\%)} & \textbf{Top 5 (\%)} \\
\midrule
\multicolumn{5}{c}{\textbf{ViM-t}} \\
\midrule
\textbf{Baseline} & \textbf{FP16} & --- & \textbf{76.07} & \textbf{93.01} \\
Uniform & W8A8 & per-channel & 75.83 & 92.85 \\
Uniform & W8A8 & per-block & 75.86 & 92.91 \\
PoT & W4A8 & per-channel & 68.18 & 88.58 \\
PoT & W4A8 & per-block & 70.31 & 89.77 \\
APoT & W4A8 & per-channel & 72.04 & 90.91 \\
\rblue \textbf{ViM-Q (Ours)} & \textbf{W4A8} & per-block & \textbf{74.23} & \textbf{92.00} \\
\midrule
\multicolumn{5}{c}{\textbf{ViM-s}} \\
\midrule
\textbf{Baseline} & \textbf{FP16} & --- & \textbf{80.48} & \textbf{95.10} \\
Uniform & W8A8 & per-channel & 80.37 & 95.00 \\
Uniform & W8A8 & per-block & 80.41 & 94.98 \\
PoT & W4A8 & per-channel & 78.04 & 93.92 \\
PoT & W4A8 & per-block & 78.78 & 94.32 \\
APoT & W4A8 & per-channel & 78.04 & 94.07 \\
\rblue \textbf{ViM-Q (Ours)} & \textbf{W4A8} & per-block & \textbf{79.89} & \textbf{94.80} \\
\midrule
\multicolumn{5}{c}{\textbf{ViM-b}} \\
\midrule
\textbf{Baseline} & \textbf{FP16} & --- & \textbf{81.88} & \textbf{95.76} \\
Uniform & W8A8 & per-channel & 81.86 & 95.75 \\
Uniform & W8A8 & per-block & 81.83 & 95.75 \\
PoT & W4A8 & per-channel & 81.09 & 95.39 \\
PoT & W4A8 & per-block & 81.35 & 95.51 \\
APoT & W4A8 & per-channel & 80.83 & 95.36 \\
\rblue \textbf{ViM-Q (Ours)} & \textbf{W4A8} & per-block & \textbf{81.67} & \textbf{95.62} \\
\bottomrule
\end{tabular}
}
\vspace{-5pt}
\end{table}

Table~\ref{tab:evaluation_sw_quantization} demonstrates the efficacy of our proposed quantization across the ViM model family.
We benchmark our approach against standard uniform quantization and constrained Power-of-Two (PoT) schemes. While APoT uses additive combinations, standard PoT is restricted to single power-of-two values.
The results highlight the impact of quantization granularity, where per-block strategies consistently outperform per-channel approaches for all models.
For ViM-t, standard 4-bit PoT quantization induces significant accuracy degradation.
In contrast, APoT recovers substantial performance, achieving 74.23\% Top-1 and 92.00\% Top-5 accuracy.
Notably, as model capacity increases, the quantization shows exceptional robustness. For ViM-s and ViM-b, our method achieves negligible accuracy loss compared to the FP16 baseline. Specifically, ViM-b shows a negligible Top-1 drop of only 0.21\%. These results confirm that \SystemName provides an optimal balance between storage efficiency and task performance.

\subsubsection{Design Space Exploration}

\begin{figure}[t]
  \centering
  \includegraphics[width=0.95\linewidth]{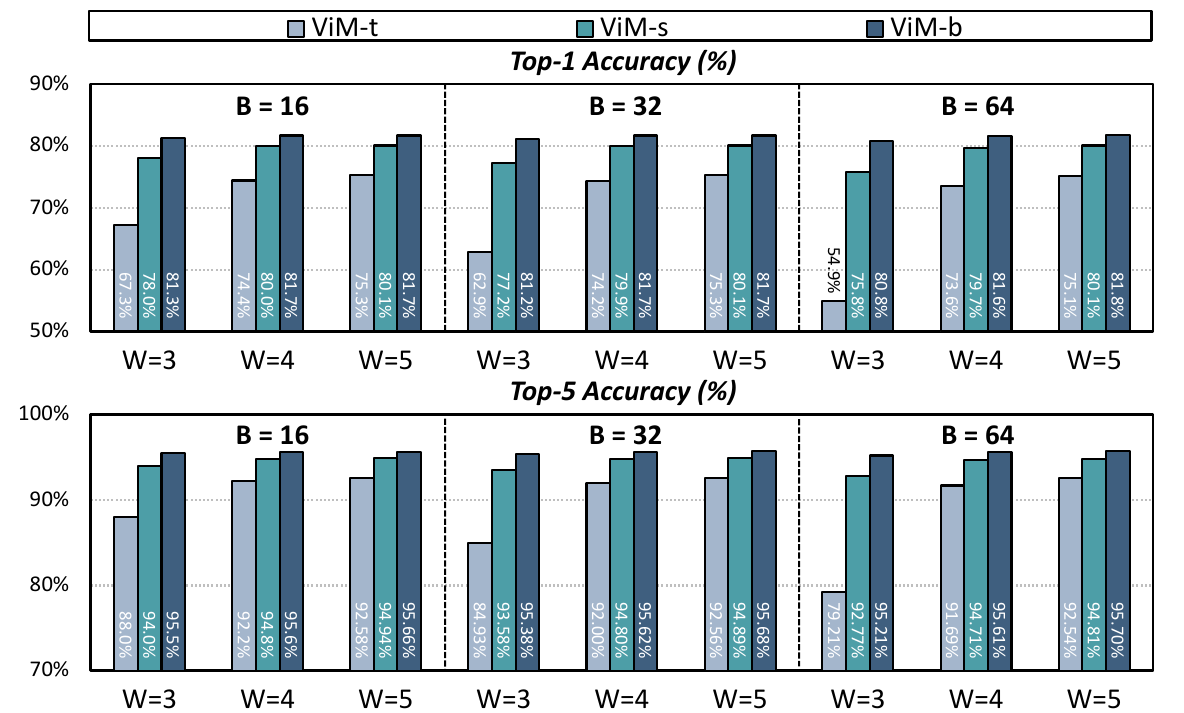}
  \captionsetup{skip=1pt}
  \caption{Design space exploration of weight bit-widths $W$ and quantization block sizes $B$.}
  \vspace{-6pt}
  \label{fig:evaluation_sw_dse}
\end{figure}

We conduct a design space exploration across weight bit-widths $W \in \{3, 4, 5\}$ and block sizes $B \in \{16, 32, 64\}$, as illustrated in Figure~\ref{fig:evaluation_sw_dse}.
Results indicate a sharp performance cliff below 4 bits: for ViM-t ($B=32$), reducing $W$ from 4 to 3 degrades Top-1 accuracy by 11.3\%, whereas increasing to $W=5$ yields diminishing returns, with the larger ViM-b exhibiting identical accuracy.
This suggests that 4-bit weights provide sufficient representational capacity without the storage overhead of 5-bit.
Regarding quantization granularity, while larger models remain robust to block size variations, ViM-t is more sensitive to coarse granularity.
While increasing to $B=64$ causes distinct degradation compared to $B=32$, refining the granularity further to $B=16$ yields negligible returns.
Consequently, we select $B=32$ as the global configuration to secure fidelity for compact models while minimizing hardware overhead.

\subsubsection{Ablation Study of Quantization Scheme}

\begin{figure}[t]
  \centering
  \includegraphics[width=0.85\linewidth]{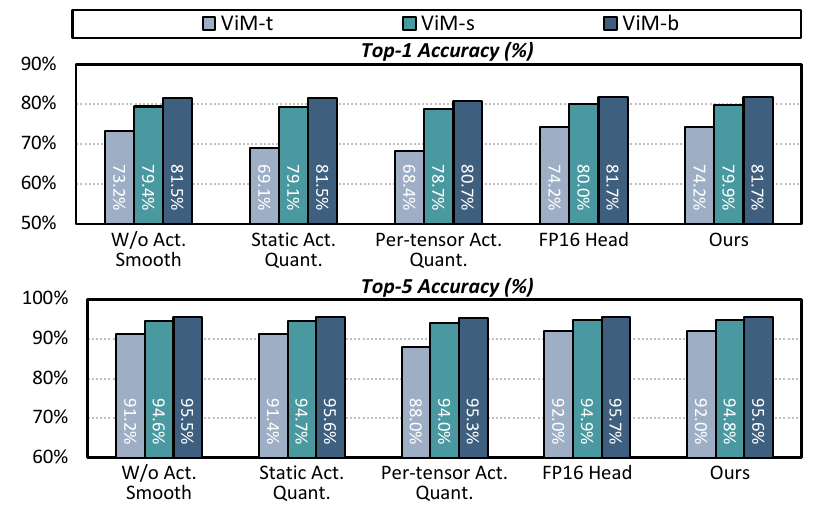}
  \captionsetup{skip=-1pt}
  \caption{Ablation studies on the key components of our activation and weight quantization framework.}
  \vspace{-5pt}
  \label{fig:evaluation_sw_ablation_study}
\end{figure}

Figure~\ref{fig:evaluation_sw_ablation_study} summarizes the ablation study validating individual quantization components.
First, removing activation smoothing causes consistent degradation across all models, with the tiny model suffering a 1.0\% drop in Top-1 accuracy, confirming its necessity for outlier mitigation.
Next, replacing our fine-grained dynamic activation quantization with a static scheme leads to a catastrophic 5.1\% drop for ViM-t, indicating static ranges fail to capture rapid distribution shifts.
Similarly, coarsening activation quantization granularity to per-tensor yields the poorest Top-1 accuracy at 68.4\%. These findings underscore that both dynamic range adjustment and fine-grained granularity for activations are critical.
Finally, the performance gap between the fully quantized and full-precision classification head is negligible. This justifies our decision to quantize all linear transformations for maximum hardware efficiency without resorting to mixed-precision overheads.
\subsection{Evaluation of Hardware Design}

\begin{table}[t]
\centering
\caption{Resource utilization of our proposed hardware implementation on ZCU102 FPGA.}
\label{tab:evaluation_hw_resources}
\vspace{-5pt}
\resizebox{0.85\columnwidth}{!}{
\begin{tabular}{l l c c}
\toprule
\textbf{Category} & \textbf{Metric} & \textbf{Utilization} & \textbf{Percentage} \\ 
\midrule
\multirow{2}{*}{\textbf{Timing}} 
& Clock Frequency & 350 MHz & -- \\
& Worst Negative Slack & 0.040 ns & -- \\ 
\midrule
\multirow{5}{*}{\textbf{Resources}}
& LUTs as Logic & 121,721 & 44.41\% \\ 
& LUTs as Memory & 27,520 & 19.11\% \\ 
& Flip-Flops & 277,278 & 50.58\% \\ 
& Block RAM Tiles & 737 & 80.81\% \\ 
& DSP48 Slices & 1,377 & 54.64\% \\ 
\midrule
\multirow{2}{*}{\textbf{Power}} 
& Dynamic Power & 11.96 W & -- \\
& Static Power & 0.81 W & -- \\
\bottomrule
\end{tabular}
}
\vspace{-5pt}
\end{table}

\subsubsection{Physical Implementation and Resource Utilization}
\begin{figure}[t]
  \centering
  \includegraphics[width=0.98\linewidth]{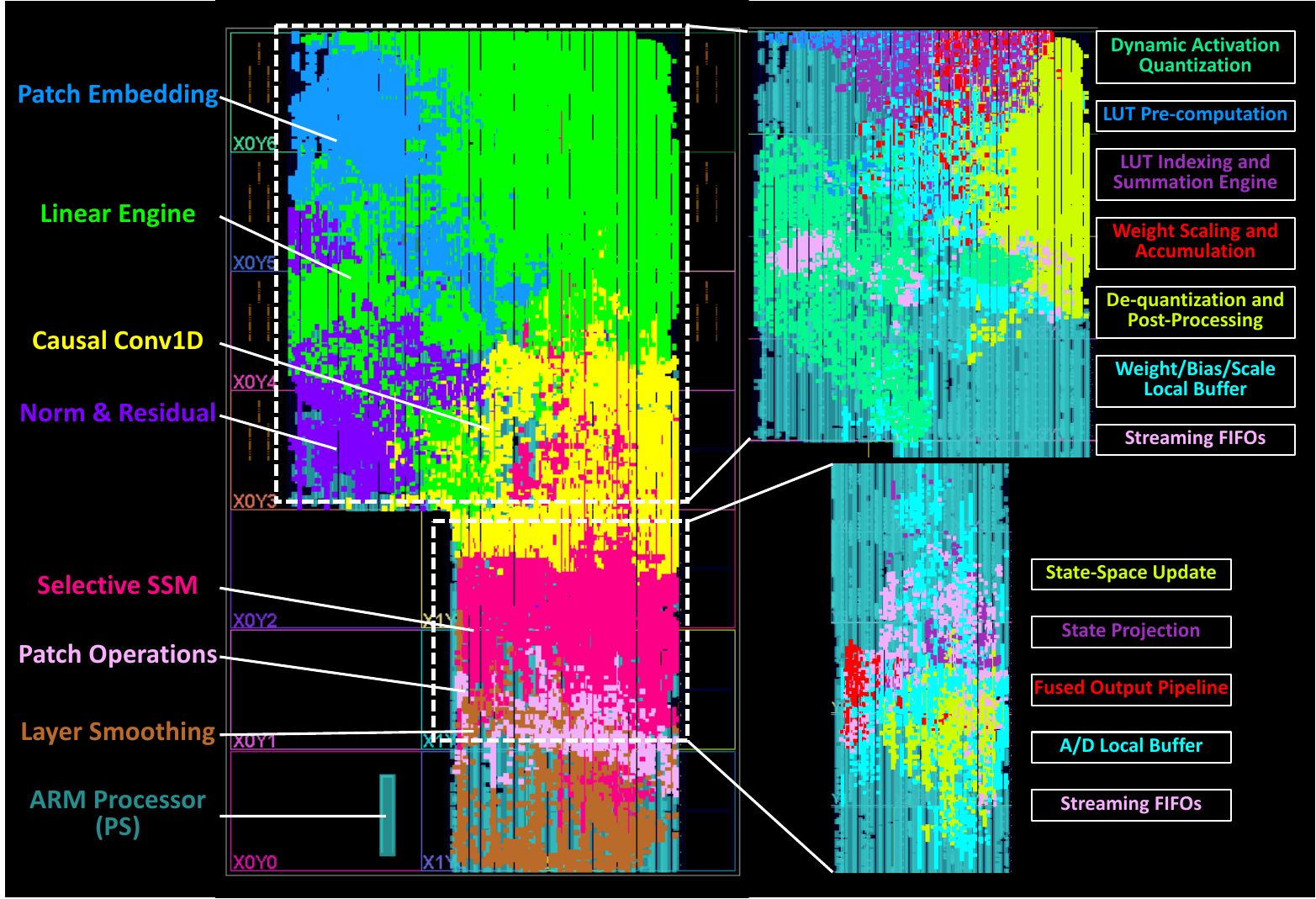}
  \captionsetup{skip=2pt}
  \caption{Physical floorplan of the implementation on ZCU102 FPGA after place-and-route.}
  \vspace{-5pt}
  \label{fig:evaluation_hw_physical_impl}
\end{figure}

We implemented the proposed \SystemName accelerator on an AMD ZCU102 FPGA. Figure~\ref{fig:evaluation_hw_physical_impl} shows the post-implementation floorplan.
Table~\ref{tab:evaluation_hw_resources} summarizes the post-route metrics. The design achieves robust timing closure at 350 MHz.
Block RAMs are the dominant resource, utilizing 80.81\% of capacity for buffering weights and intermediate states to minimize off-chip access. Logic (LUTs/FFs) and DSP usage remain balanced at approximately 50\%, leaving room for expansion. Total power consumption is approximately 12.77 W.

\begin{table*}[t]
\centering
\caption{Ablation study of the unified W4A8 linear engine. We evaluate the impact of incremental optimizations on single-layer latency (In=192, Out=384) and synthesis resource utilization. Speedup is normalized to the spatial tiling stage.}
\label{tab:linear_layer_eval}
\vspace{-5pt}
\resizebox{0.78\linewidth}{!}{
\begin{tabular}{l c c c c c c}
\toprule
\multirow{2}{*}{\textbf{Optimization Stages}} & 
\multicolumn{2}{c}{\textbf{Single Layer (In=192, Out=384)}} &
\multicolumn{4}{c}{\textbf{Linear Engine Utilization}} \\
\cmidrule(lr){2-3} \cmidrule(lr){4-7}
& Latency (cycles) & Speedup & BRAM & DSP & FF & LUT \\
\midrule
\textbf{Baseline}: W4A8 GEMM (Sequential Scalar) & 
    7,435,649 & 0.02$\times$ & 
    281 & 46 & 25,738 & 32,378 \\
\midrule
\hspace{3mm}+ Spatial Tiling \& Parallel Adder Trees & 
    147,780 & 1.00$\times$ & 
    539 & 329 & 167,757 & 161,587 \\
\hspace{3mm}+ Memory-Aligned Weight Reordering & 
    58,780 & 2.51$\times$ & 
    539 & 326 & 137,177 & 102,819 \\
\hspace{3mm}+ Hardware-Efficient Decoding via Bit-Shifting & 
    58,840 & 2.51$\times$ & 
    539 & 326 & 131,100 & 77,070 \\
\hspace{3mm}+ Precision-Preserving Activation Pre-Shifting & 
    58,780 & 2.51$\times$ & 
    539 & 326 & 132,932 & 94,336 \\
\rowcolor{gray!10}
\hspace{3mm}\textbf{+ LUT-Based APoT Precomputation (Ours)} & 
    \textbf{58,780} & \textbf{2.51$\times$} & 
    \textbf{539} & \textbf{326} & \textbf{142,200} & \textbf{77,918} \\
\bottomrule
\end{tabular}
}
\vspace{-10pt}
\end{table*}

\subsubsection{Performance Comparison with GPU Baselines}
\begin{table}[t]
\centering
\caption{End-to-end performance comparison of our FPGA implementation versus GPU baselines at $224 \times 224$ resolution.}
\label{tab:evaluation_hw_gpu}
\vspace{-5pt}
\resizebox{\columnwidth}{!}{
\begin{threeparttable}
\begin{tabular}{l cc cc cc}
\toprule
\multirow{2}{*}{\textbf{Metric}} &
  \multicolumn{2}{c}{\textbf{FPGA (Ours)}} &
  \multicolumn{2}{c}{\textbf{GPU (Quant.)}} &
  \multicolumn{2}{c}{\textbf{GPU (FP16)}} \\
\cmidrule(lr){2-3} \cmidrule(lr){4-5} \cmidrule(l){6-7}
& ViM-t & ViM-s & ViM-t & ViM-s & ViM-t & ViM-s \\
\midrule
Frequency (MHz)
    & \multicolumn{2}{c}{350}
    & \multicolumn{2}{c}{1695}
    & \multicolumn{2}{c}{1695} \\
\midrule
Process (nm)
    & \multicolumn{2}{c}{16}
    & \multicolumn{2}{c}{8}
    & \multicolumn{2}{c}{8} \\
\midrule
Precision
    & \multicolumn{2}{c}{W4A8}
    & \multicolumn{2}{c}{W4A8}
    & \multicolumn{2}{c}{FP16} \\
\midrule
Latency (ms)
    & 53.76 & 127.3 & 117.1 & 119.1 & 54.60 & 60.86 \\
Power (W)\textsuperscript{$\dagger$}
    & 12.77 & 12.77 & 153.7 & 155.7 & 201.1 & 205.9 \\
Energy (J)
    & 0.686 & 1.625 & 18.00 & 18.54 & 10.98 & 12.53 \\
\midrule
Power Efficiency
    & 1.00 & 1.00 & 0.038$ \times$ & 0.088$ \times$ & 0.063$ \times$ & 0.130$ \times$ \\
\bottomrule
\end{tabular}
\begin{tablenotes}
\scriptsize
\item[$\dagger$] GPU power excludes estimated 50\,W static board and fan power.
\end{tablenotes}
\end{threeparttable}
}
\vspace{-3pt}
\end{table}

We benchmark end-to-end inference for ViM-t and ViM-s against GPU baselines in Table~\ref{tab:evaluation_hw_gpu}. The results expose the fundamental structural limitations of GPU architectures in handling low-batch inference and dynamic quantization workloads.

A critical finding is that the quantized GPU baseline is significantly slower than its full-precision counterpart, lagging by over $2\times$ for ViM-t.
This highlights the high runtime overhead of dynamic quantization on GPUs, as calculating per-token scaling factors negates the speedup of INT8 arithmetic. In contrast, our FPGA architecture integrates the dynamic quantizer directly into the streaming pipeline. This allows us to outperform the quantized GPU on latency across all tests and surpass the FP16 GPU on the compact ViM-t model.

Despite utilizing an older 16~nm process compared to the GPU's 8~nm node, our accelerator demonstrates decisive superiority in efficiency. By eliminating the power-hungry control logic required for GPU flexibility, we achieve a $26.3\times$ energy efficiency gain over the quantized GPU baseline for ViM-t, and an $11.4\times$ advantage for ViM-s.

\subsubsection{Throughput and Energy Scaling}
\begin{figure}[t]
    \centering
    \vspace{-5pt}
    \includegraphics[width=\linewidth]{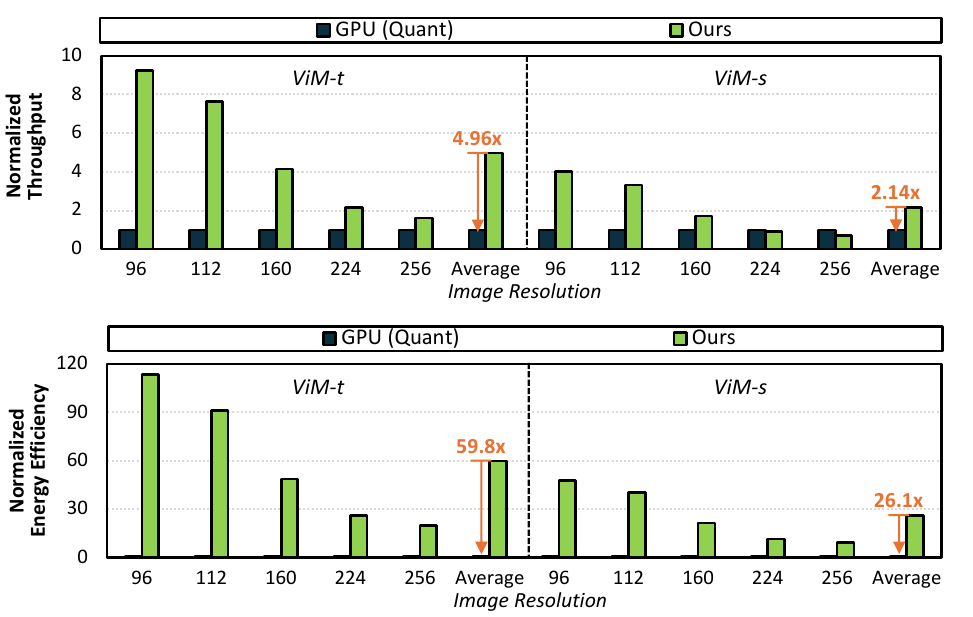}
    \captionsetup{skip=-1pt}
    \caption{Normalized throughput and energy efficiency across varying input resolutions. The GPU baseline is normalized to one.}
    \vspace{-5pt}
    \label{fig:evaluation_hw_normalized_energy}
\end{figure}

Figure~\ref{fig:evaluation_hw_normalized_energy} evaluates the scalability of throughput and energy efficiency across varying input resolutions. The results highlight the severe inefficiency of GPUs when handling the low-workload scenarios typical of edge deployment. 
At low resolutions, the GPU suffers from compute under-utilization, as the short sequence lengths fail to saturate its massive parallel cores. In contrast, our FPGA pipeline operates with minimal startup latency, delivering immediate throughput. This results in a peak acceleration of over $9\times$ at $96 \times 96$, with an average speedup of $4.96\times$ for ViM-t and $2.14\times$ for ViM-s across all resolutions relative to the quantized GPU.
Furthermore, \SystemName demonstrates exceptional energy efficiency. Even as the resolution increases and the GPU begins to amortize its overhead, our design maintains a decisive advantage. Overall, we achieve an average energy efficiency gain of $59.8\times$ for ViM-t and $26.1\times$ for ViM-s.

\subsubsection{Ablation Study of the Linear Engine Design}

Table~\ref{tab:linear_layer_eval} details our incremental linear engine optimizations, which focus on maximizing throughput and minimizing logic overhead.
First, although spatial tiling unrolls the computation to achieve significant acceleration, it creates irregular, non-sequential memory access patterns.
We resolve this via memory-aligned weight reordering, ensuring purely sequential fetches that minimize latency cycles and simplify control logic.
Next, by leveraging APoT properties, we replace standard multipliers with lightweight bit-shifts, further dropping LUT consumption to roughly 77\,K.
Finally, while precision-preserving activation pre-shifting is necessary for accuracy, it introduces logic overhead. We neutralize this through LUT-based APoT precomputation.
By pre-computing APoT terms once per input tile rather than shifting within PEs, we remove complexity from the critical path, yielding a highly efficient final footprint of 78\,K LUTs.
\section{Conclusions}
\label{sec:conclusion}

We present \SystemName, an algorithm-hardware co-design framework enabling efficient, end-to-end Vision Mamba inference on the edge. By integrating a novel quantization scheme with a runtime-parameterizable FPGA accelerator featuring a unified LUT-based linear engine and a fine-grained pipelined SSM engine, our implementation achieves lower latency and superior energy efficiency compared to GPU baselines.
\section*{Acknowledgment}
This work was supported by the Research Grants Council of the Hong Kong Special Administrative Region, China (Project No. CityU 11202925, CityU 11202124, and CityU 11201422), the Innovation and Technology Commission of Hong Kong (Project No. MHP/072/23), and City University of Hong Kong (Project 9678187).

\balance
\bibliographystyle{IEEEtran}
\bibliography{citations}

@article{attention,
  title={Attention is all you need},
  author={Vaswani, Ashish and Shazeer, Noam and Parmar, Niki and Uszkoreit, Jakob and Jones, Llion and Gomez, Aidan N and Kaiser, {\L}ukasz and Polosukhin, Illia},
  journal={Advances in Neural Information Processing Systems (NeurIPS)},
  volume={30},
  year={2017}
}

@inproceedings{vit,
  title={An Image is Worth 16x16 Words: Transformers for Image Recognition at Scale},
  author={Alexey Dosovitskiy and Lucas Beyer and Alexander Kolesnikov and Dirk Weissenborn and Xiaohua Zhai and Thomas Unterthiner and Mostafa Dehghani and Matthias Minderer and Georg Heigold and Sylvain Gelly and Jakob Uszkoreit and Neil Houlsby},
  booktitle={Proceedings of the International Conference on Learning Representations (ICLR)},
  year={2021}
}

@inproceedings{swin,
  title={Swin transformer: Hierarchical vision transformer using shifted windows},
  author={Liu, Ze and Lin, Yutong and Cao, Yue and Hu, Han and Wei, Yixuan and Zhang, Zheng and Lin, Stephen and Guo, Baining},
  booktitle={Proceedings of the IEEE/CVF International Conference on Computer Vision (ICCV)},
  pages={10012--10022},
  year={2021}
}

@inproceedings{deit,
  title={Training data-efficient image transformers \& distillation through attention},
  author={Touvron, Hugo and Cord, Matthieu and Douze, Matthijs and Massa, Francisco and Sablayrolles, Alexandre and J{\'e}gou, Herv{\'e}},
  booktitle={Proceedings of the International Conference on Machine Learning (ICML)},
  pages={10347--10357},
  year={2021}
}

@inproceedings{mae,
  title={Masked autoencoders are scalable vision learners},
  author={He, Kaiming and Chen, Xinlei and Xie, Saining and Li, Yanghao and Doll{\'a}r, Piotr and Girshick, Ross},
  booktitle={Proceedings of the IEEE/CVF Conference on Computer Vision and Pattern Recognition (CVPR)},
  pages={16000--16009},
  year={2022}
}

@article{dynamicvit,
  title={Dynamicvit: Efficient vision transformers with dynamic token sparsification},
  author={Rao, Yongming and Zhao, Wenliang and Liu, Benlin and Lu, Jiwen and Zhou, Jie and Hsieh, Cho-Jui},
  journal={Advances in Neural Information Processing Systems (NeurIPS)},
  volume={34},
  pages={13937--13949},
  year={2021}
}

@inproceedings{beit,
  title={{BE}iT: {BERT} Pre-Training of Image Transformers},
  author={Hangbo Bao and Li Dong and Songhao Piao and Furu Wei},
  booktitle={Proceedings of the International Conference on Learning Representations (ICLR)},
  year={2022}
}

@inproceedings{taming,
  title={Taming transformers for high-resolution image synthesis},
  author={Esser, Patrick and Rombach, Robin and Ommer, Bjorn},
  booktitle={Proceedings of the IEEE/CVF Conference on Computer Vision and Pattern Recognition (CVPR)},
  pages={12873--12883},
  year={2021}
}

@article{vitsurvey1,
  title={Transformers in vision: A survey},
  author={Khan, Salman and Naseer, Muzammal and Hayat, Munawar and Zamir, Syed Waqas and Khan, Fahad Shahbaz and Shah, Mubarak},
  journal={ACM Computing Surveys (CSUR)},
  volume={54},
  number={10s},
  pages={1--41},
  year={2022}
}

@article{vitsurvey2,
  title={A survey on vision transformer},
  author={Han, Kai and Wang, Yunhe and Chen, Hanting and Chen, Xinghao and Guo, Jianyuan and Liu, Zhenhua and Tang, Yehui and Xiao, An and Xu, Chunjing and Xu, Yixing and others},
  journal={IEEE Transactions on Pattern Analysis and Machine Intelligence (TPAMI)},
  volume={45},
  number={1},
  pages={87--110},
  year={2022}
}

@inproceedings{mamba,
  title={Mamba: Linear-time sequence modeling with selective state spaces},
  author={Gu, Albert and Dao, Tri},
  booktitle={Conference on Language Modeling (COLM)},
  year={2024}
}

@inproceedings{mamba2,
  title={Transformers are SSMs: generalized models and efficient algorithms through structured state space duality},
  author={Dao, Tri and Gu, Albert},
  booktitle={Proceedings of the International Conference on Machine Learning (ICML)},
  pages={10041--10071},
  year={2024}
}

@inproceedings{vim,
  title={Vision Mamba: Efficient Visual Representation Learning with Bidirectional State Space Model},
  author={Zhu, Lianghui and Liao, Bencheng and Zhang, Qian and Wang, Xinlong and Liu, Wenyu and Wang, Xinggang},
  booktitle={Proceedings of the International Conference on Machine Learning (ICML)},
  year={2024}
}

@article{vmamba,
  title={Vmamba: Visual state space model},
  author={Liu, Yue and Tian, Yunjie and Zhao, Yuzhong and Yu, Hongtian and Xie, Lingxi and Wang, Yaowei and Ye, Qixiang and Jiao, Jianbin and Liu, Yunfan},
  journal={Advances in Neural Information Processing Systems (NeurIPS)},
  volume={37},
  pages={103031--103063},
  year={2024}
}

@inproceedings{mambaout,
  title={Mambaout: Do we really need mamba for vision?},
  author={Yu, Weihao and Wang, Xinchao},
  booktitle={Proceedings of the IEEE/CVF Conference on Computer Vision and Pattern Recognition (CVPR)},
  pages={4484--4496},
  year={2025}
}

@inproceedings{mambavision,
  title={Mambavision: A hybrid mamba-transformer vision backbone},
  author={Hatamizadeh, Ali and Kautz, Jan},
  booktitle={Proceedings of the IEEE/CVF Conference on Computer Vision and Pattern Recognition (CVPR)},
  pages={25261--25270},
  year={2025}
}

@inproceedings{mobilemamba,
  title={Mobilemamba: Lightweight multi-receptive visual mamba network},
  author={He, Haoyang and Zhang, Jiangning and Cai, Yuxuan and Chen, Hongxu and Hu, Xiaobin and Gan, Zhenye and Wang, Yabiao and Wang, Chengjie and Wu, Yunsheng and Xie, Lei},
  booktitle={Proceedings of the IEEE/CVF Conference on Computer Vision and Pattern Recognition (CVPR)},
  pages={4497--4507},
  year={2025}
}

@inproceedings{localmamba,
  title={Localmamba: Visual state space model with windowed selective scan},
  author={Huang, Tao and Pei, Xiaohuan and You, Shan and Wang, Fei and Qian, Chen and Xu, Chang},
  booktitle={Proceedings of the European Conference on Computer Vision (ECCV)},
  pages={12--22},
  year={2024}
}

@article{vimsurvey1,
  title={Mamba in vision: A comprehensive survey of techniques and applications},
  author={Rahman, Md Maklachur and Tutul, Abdullah Aman and Nath, Ankur and Laishram, Lamyanba and Jung, Soon Ki and Hammond, Tracy},
  journal={arXiv preprint arXiv:2410.03105 (arXiv)},
  year={2024}
}

@article{vimsurvey2,
  title={Vision mamba: A comprehensive survey and taxonomy},
  author={Liu, Xiao and Zhang, Chenxu and Huang, Fuxiang and Xia, Shuyin and Wang, Guoyin and Zhang, Lei},
  journal={IEEE Transactions on Neural Networks and Learning Systems (TNNLS)},
  year={2025}
}

@article{vimsurvey3,
  title={A survey on mamba architecture for vision applications},
  author={Ibrahim, Fady and Liu, Guangjun and Wang, Guanghui},
  journal={arXiv preprint arXiv:2502.07161 (arXiv)},
  year={2025}
}

@inproceedings{ptq4vm,
  title={PTQ4VM: Post-training quantization for visual mamba},
  author={Cho, Younghyun and Lee, Changhun and Kim, Seonggon and Park, Eunhyeok},
  booktitle={Proceedings of the IEEE/CVF Winter Conference on Applications of Computer Vision (WACV)},
  pages={1176--1185},
  year={2025}
}

@article{mambaptq,
  title={Mamba-ptq: Outlier channels in recurrent large language models},
  author={Pierro, Alessandro and Abreu, Steven},
  journal={arXiv preprint arXiv:2407.12397 (arXiv)},
  year={2024}
}

@inproceedings{mambaquant,
  title={MambaQuant: Quantizing the Mamba Family with Variance Aligned Rotation Methods},
  author={Zukang Xu and Yuxuan Yue and Xing Hu and Dawei Yang and Zhihang Yuan and Zixu Jiang and Zhixuan Chen and JiangyongYu and XUCHEN and Sifan Zhou},
  booktitle={Proceedings of the International Conference on Learning Representations (ICLR)},
  year={2025}
}

@inproceedings{quamba,
  title = {Quamba: A Post-Training Quantization Recipe for Selective State Space Models},
  author = {Chiang, Hung-Yueh and Chang, Chi-Chih and Frumkin, Natalia and Wu, Kai-Chiang and Marculescu, Diana},
  booktitle = {Proceedings of the International Conference on Learning Representations (ICLR)},
  year = {2025}
}

@inproceedings{quamba2,
  title = {Quamba2: A Robust and Scalable Post-training Quantization Framework for Selective State Space Models},
  author = {Chiang, Hung-Yueh and Chang, Chi-Chih and Frumkin, Natalia and Wu, Kai-Chiang and Abdelfattah, Mohamed S and Marculescu, Diana},
  booktitle = {Proceedings of the International Conference on Machine Learning (ICML)},
  year = {2025}
}

@INPROCEEDINGS{mambakscale,
  title={Post-Training Quantization for Vision Mamba with K-Scaled Quantization and Reparameterization}, 
  author={Shi, Bo-Yun and Lo, Yi-Cheng and Wu, An-Yeu and Tsai, Yi-Min},
  booktitle={Proceedings of the International Workshop on Machine Learning for Signal Processing (MLSP)}, 
  year={2025},
  pages={1-6}
}

@inproceedings{vimvq,
  title={ViM-VQ: Efficient Post-Training Vector Quantization for Visual Mamba},
  author={Deng, Juncan and Li, Shuaiting and Wang, Zeyu and Xu, Kedong and Gu, Hong and Huang, Kejie},
  booktitle={Proceedings of the IEEE/CVF International Conference on Computer Vision (ICCV)},
  pages={24518--24527},
  year={2025}
}

@inproceedings{ouromamba,
  title={OuroMamba: A Data-Free Quantization Framework for Vision Mamba},
  author={Ramachandran, Akshat and Lee, Mingyu and Xu, Huan and Kundu, Souvik and Krishna, Tushar},
  booktitle={Proceedings of the IEEE/CVF International Conference on Computer Vision (ICCV)},
  pages={21177--21186},
  year={2025}
}

@inproceedings{marca,
  title={Marca: Mamba accelerator with reconfigurable architecture},
  author={Li, Jinhao and Huang, Shan and Xu, Jiaming and Liu, Jun and Ding, Li and Xu, Ningyi and Dai, Guohao},
  booktitle={Proceedings of the IEEE/ACM International Conference on Computer-Aided Design (ICCAD)},
  pages={1--9},
  year={2024}
}

@inproceedings{lightmamba,
  title={Lightmamba: Efficient mamba acceleration on fpga with quantization and hardware co-design},
  author={Wei, Renjie and Xu, Songqiang and Zhong, Linfeng and Yang, Zebin and Guo, Qingyu and Wang, Yuan and Wang, Runsheng and Li, Meng},
  booktitle={Proceedings of the Conference on Design, Automation and Test in Europe (DATE)},
  pages={1--7},
  year={2025}
}

@inproceedings{mamba2fpga,
  title={An Efficient FPGA-Based Hardware Accelerator of Fully Quantized Mamba-2},
  author={Zhou, Kailing and Jiao, Han and Huang, Wenjin and Huang, Yihua},
  booktitle={Proceedings of the IEEE International Symposium on Field-Programmable Custom Computing Machines (FCCM)},
  pages={217--226},
  year={2025}
}

@inproceedings{mambax,
  title={Mamba-X: An End-to-End Vision Mamba Accelerator for Edge Computing Devices},
  author={Yoon, Dongho and Lee, Gungyu and Chang, Jaewon and Lee, Yunjae and Lee, Dongjae and Rhu, Minsoo},
  booktitle={Proceedings of the IEEE/ACM International Conference on Computer-Aided Design (ICCAD)},
  pages={1--9},
  year={2025}
}

@inproceedings{specmamba,
  title={SpecMamba: Accelerating Mamba Inference on FPGA with Speculative Decoding},
  author={Zhong, Linfeng and Xu, Songqiang and Wen, Huifeng and Xie, Tong and Guo, Qingyu and Wang, Yuan and Li, Meng},
  booktitle={Proceedings of the IEEE/ACM International Conference on Computer-Aided Design (ICCAD)},
  pages={1--9},
  year={2025},
  organization={IEEE}
}

@inproceedings{fastmamba,
  title={Fastmamba: A high-speed and efficient mamba accelerator on fpga with accurate quantization},
  author={Wang, Aotao and Shao, Haikuo and Ma, Shaobo and Wang, Zhongfeng},
  booktitle={Proceedings of the IEEE Computer Society Annual Symposium on VLSI (ISVLSI)},
  volume={1},
  pages={1--6},
  year={2025}
}

@inproceedings{mixmatch,
  title={Mix and match: A novel fpga-centric deep neural network quantization framework},
  author={Chang, Sung-En and Li, Yanyu and Sun, Mengshu and Shi, Runbin and So, Hayden K-H and Qian, Xuehai and Wang, Yanzhi and Lin, Xue},
  booktitle={Proceedings of the IEEE International Symposium on High-Performance Computer Architecture (HPCA)},
  pages={208--220},
  year={2021}
}

@article{p2vit,
  title={P$^2$-ViT: Power-of-Two Post-Training Quantization and Acceleration for Fully Quantized Vision Transformer},
  author={Shi, Huihong and Cheng, Xin and Mao, Wendong and Wang, Zhongfeng},
  journal={IEEE Transactions on Very Large Scale Integration Systems (TVLSI)},
  year={2024}
}

@inproceedings{li2020Additive,
  title={Additive Powers-of-Two Quantization: An Efficient Non-uniform Discretization for Neural Networks},
  author={Yuhang Li and Xin Dong and Wei Wang},
  booktitle={Proceedings of the International Conference on Learning Representations (ICLR)},
  year={2020}
}

@inproceedings{yao2022rapq,
  title = {RAPQ: Rescuing Accuracy for Power-of-Two Low-bit Post-training Quantization},
  author = {Yao, Hongyi and Li, Pu and Cao, Jian and Liu, Xiangcheng and Xie, Chenying and Wang, Bingzhang},
  booktitle = {Proceedings of the International Joint Conference on Artificial Intelligence (IJCAI)},
  pages = {1573--1579},
  year = {2022},
}

@article{przewlocka2022power,
  title={Power-of-two quantization for low bitwidth and hardware compliant neural networks},
  author={Przewlocka-Rus, Dominika and Sarwar, Syed Shakib and Sumbul, H Ekin and Li, Yuecheng and De Salvo, Barbara},
  journal={arXiv preprint arXiv:2203.05025 (arXiv)},
  year={2022}
}

@inproceedings{smoothquant,
  title={Smoothquant: Accurate and efficient post-training quantization for large language models},
  author={Xiao, Guangxuan and Lin, Ji and Seznec, Mickael and Wu, Hao and Demouth, Julien and Han, Song},
  booktitle={Proceedings of the International Conference on Machine Learning (ICML)},
  pages={38087--38099},
  year={2023}
}

@inproceedings{svdquant,
  title={{SVDQ}uant: Absorbing Outliers by Low-Rank Component for 4-Bit Diffusion Models},
  author={Muyang Li and Yujun Lin and Zhekai Zhang and Tianle Cai and Junxian Guo and Xiuyu Li and Enze Xie and Chenlin Meng and Jun-Yan Zhu and Song Han},
  booktitle={Proceedings of the International Conference on Learning Representations (ICLR)},
  year={2025}
}

@article{vsquant,
  title={Vs-quant: Per-vector scaled quantization for accurate low-precision neural network inference},
  author={Dai, Steve and Venkatesan, Rangha and Ren, Mark and Zimmer, Brian and Dally, William and Khailany, Brucek},
  journal={Machine Learning and Systems (MLSys)},
  volume={3},
  pages={873--884},
  year={2021}
}

@article{zeroquant,
  title={Zeroquant: Efficient and affordable post-training quantization for large-scale transformers},
  author={Yao, Zhewei and Yazdani Aminabadi, Reza and Zhang, Minjia and Wu, Xiaoxia and Li, Conglong and He, Yuxiong},
  journal={Advances in Neural Information Processing Systems (NeurIPS)},
  volume={35},
  pages={27168--27183},
  year={2022}
}

@article{gpt3,
  title={LLM.int8(): 8-bit matrix multiplication for transformers at scale},
  author={Dettmers, Tim and Lewis, Mike and Belkada, Younes and Zettlemoyer, Luke},
  journal={Advances in Neural Information Processing Systems (NeurIPS)},
  volume={35},
  pages={30318--30332},
  year={2022}
}

@inproceedings{scan,
  title={Prefix sums and their applications},
  author={Guy E. Blelloch},
  year={1990}
}

@inproceedings{edgemoe,
  title={Edge-moe: Memory-efficient multi-task vision transformer architecture with task-level sparsity via mixture-of-experts},
  author={Sarkar, Rishov and Liang, Hanxue and Fan, Zhiwen and Wang, Zhangyang and Hao, Cong},
  booktitle={Proceedings of the IEEE/ACM International Conference on Computer Aided Design (ICCAD)},
  pages={01--09},
  year={2023}
}

@inproceedings{finn,
  title={Finn: A framework for fast, scalable binarized neural network inference},
  author={Umuroglu, Yaman and Fraser, Nicholas J and Gambardella, Giulio and Blott, Michaela and Leong, Philip and Jahre, Magnus and Vissers, Kees},
  booktitle={Proceedings of the ACM/SIGDA International Symposium on Field-Programmable Gate Arrays (ISFPGA)},
  pages={65--74},
  year={2017}
}

@inproceedings{famous,
  title={FAMOUS: Flexible accelerator for the attention mechanism of transformer on UltraScale+ FPGAs},
  author={Kabir, Ehsan and Kabir, Md Arafat and Downey, Austin RJ and Bakos, Jason D and Andrews, David and Huang, Miaoqing},
  booktitle={Proceedings of the International Conference on Field Programmable Technology (ICFPT)},
  pages={1--2},
  year={2024}
}

@article{awq,
  title={Awq: Activation-aware weight quantization for on-device llm compression and acceleration},
  author={Lin, Ji and Tang, Jiaming and Tang, Haotian and Yang, Shang and Chen, Wei-Ming and Wang, Wei-Chen and Xiao, Guangxuan and Dang, Xingyu and Gan, Chuang and Han, Song},
  journal={Machine Learning and Systems (MLSys)},
  volume={6},
  pages={87--100},
  year={2024}
}

@inproceedings{ladder,
  title={Ladder: Enabling efficient Low-Precision deep learning computing through hardware-aware tensor transformation},
  author={Wang, Lei and Ma, Lingxiao and Cao, Shijie and Zhang, Quanlu and Xue, Jilong and Shi, Yining and Zheng, Ningxin and Miao, Ziming and Yang, Fan and Cao, Ting and others},
  booktitle={Proceedings of the USENIX Symposium on Operating Systems Design and Implementation (OSDI)},
  pages={307--323},
  year={2024}
}

@inproceedings{lut,
  title={LUT Tensor Core: A Software-Hardware Co-Design for LUT-Based Low-Bit LLM Inference},
  author={Mo, Zhiwen and Wang, Lei and Wei, Jianyu and Zeng, Zhichen and Cao, Shijie and Ma, Lingxiao and Jing, Naifeng and Cao, Ting and Xue, Jilong and Yang, Fan and others},
  booktitle={Proceedings of the International Symposium on Computer Architecture (ISCA)},
  pages={514--528},
  year={2025}
}

@inproceedings{tmac,
  title={T-mac: Cpu renaissance via table lookup for low-bit llm deployment on edge},
  author={Wei, Jianyu and Cao, Shijie and Cao, Ting and Ma, Lingxiao and Wang, Lei and Zhang, Yanyong and Yang, Mao},
  booktitle={Proceedings of the European Conference on Computer Systems (EuroSys)},
  pages={278--292},
  year={2025}
}

@inproceedings{lutdla,
  title={LUT-DLA: Lookup Table as Efficient Extreme Low-Bit Deep Learning Accelerator},
  author={Li, Guoyu and Ye, Shengyu and Chen, Chunyun and Wang, Yang and Yang, Fan and Cao, Ting and Liu, Cheng and Aly, Mohamed M Sabry and Yang, Mao},
  booktitle={Proceedings of the IEEE International Symposium on High Performance Computer Architecture (HPCA)},
  pages={671--684},
  year={2025},
  organization={IEEE}
}

@inproceedings{figlut,
  title={FIGLUT: An Energy-Efficient Accelerator Design for FP-INT GEMM Using Look-Up Tables},
  author={Park, Gunho and Kwon, Hyeokjun and Kim, Jiwoo and Bae, Jeongin and Park, Baeseong and Lee, Dongsoo and Lee, Youngjoo},
  booktitle={Proceedings of the IEEE International Symposium on High Performance Computer Architecture (HPCA)},
  pages={1098--1111},
  year={2025},
  organization={IEEE}
}

@inproceedings{lutgemm,
  title={LUT-GEMM: Quantized Matrix Multiplication based on LUTs for Efficient Inference in Large-Scale Generative Language Models},
  author={Gunho Park and Baeseong Park and Minsub Kim and Sungjae Lee and Jeonghoon Kim and Beomseok Kwon and Se Jung Kwon and Byeongwook Kim and Youngjoo Lee and Dongsoo Lee},
  booktitle={Proceedings of the International Conference on Learning Representations (ICLR)},
  year={2022}
}

\end{document}